\documentstyle[amssymb,amstex,multicol,epsfig,prb,aps]{revtex}

\newcommand{\gsim}{\raisebox{-0.5ex} {$\stackrel{>} {\sim}$}}
\newcommand{\lsim}{\raisebox{-0.5ex}{$\stackrel{<}{\sim}$}}

\begin{document}
\draft

\title{Monte Carlo simulation of subsurface ordering\\kinetics in
      an fcc--alloy model}

\author{M. Kessler, W. Dieterich}

\address{Fachbereich Physik,
Universit\"{a}t Konstanz, D-78457 Konstanz, Germany}

\author{and A. Majhofer}

\address{Institute of Experimental Physics, Warsaw University,
ul. Ho$\dot{z}$a 69,\\ PL-00681 Warszawa, Poland}

\date{March 1, 2001}

\maketitle

\begin{abstract}
Within the atom--vacancy exchange mechanism in a nearest--neighbor
interaction model we investigate the kinetics of surface--induced
ordering processes close to the (001) surface of an fcc $A_{3}B$--alloy. 
After a sudden quench into the ordered phase with a
final temperature above the ordering spinodal, $T_{f}>T_{sp}$, the
early time kinetics is dominated by a segregation front which
propagates into the bulk with nearly constant velocity. Below the
spinodal, $T_{f}<T_{sp}$, motion of the segregation wave reflects a
coarsening process which appears to be slower than predicted by the
Lifschitz--Allen--Cahn law. In addition, in the front--penetrated
region lateral growth differs distinctly from perpendicular growth, as
a result of the special structure of antiphase boundaries near the surface.
Our results are compared with recent experiments on the subsurface ordering 
kinetics at $\mathrm{Cu_3Au}$ (001).
\end{abstract}


\begin{multicols}{2}\narrowtext
\section*{1. Introduction}

Near a surface, kinetic processes connected with first--order phase
transitions are generally modified in a fundamental manner relative to
the corresponding bulk process. Well--known examples are
surface--induced spinodal decomposition and nucleation 
phenomena,\cite{Pu97,Fi98,Bi87,Ca89,Li86,Bl95} whose understanding 
and control is of major concern in modern
thin film and nanostructure technologies. Surface kinetic effects
during ordering transitions in metallic binary alloys \cite{Schweika,Dosch} 
acquire special interest from the scientific point of view as their
(low--temperature) ordered phase normally is characterized by a
multicomponent order parameter, whose components in general couple
differently to a crystal surface with a particular symmetry and hence
will display different relaxational behaviors. In such cases, the
surface is expected to induce rich dynamical behavior of the order
parameter, predominantly at short times. 

A well--studied system is the
(001) surface of $\mathrm{Cu_{3}Au}$, an fcc--alloy which undergoes a
first--order bulk transition from the disordered phase to the ordered
$L1_{2}$ structure at a temperature $T_{0}=663K$. In the bulk,
$\mathrm{Au}$--atoms preferentially occupy one of the four
simple--cubic sublattices of the underlying fcc--lattice. The ground
state therefore is $4$--fold degenerate, and there exist two types of
antiphase boundaries separating the $4$ types of energetically
equivalent equilibrium domains. Several peculiar modifications of
order near the (001) surface of $\mathrm{Cu_{3}Au}$ have been reported.
Below the ordering temperature $T_{0}$ the (001)--surface displays 
disorder--wetting.\cite{Do88,Di88,Li83} 
Moreover, above $T_{0}$, the tendency of Au--atoms to enrich
in the outermost (first) layer leads to an
oscillatory segregation profile with successive Au--depletion and
enrichment in even and odd layers, 
respectively.\cite{Rei95,Me95,Hay98,Se97} This profile decays
towards the bulk on a length scale given by the bulk--correlation
length $\xi(T)$, which on extrapolation to lower temperatures appears
to diverge at the spinodal temperature 
$T_{sp}\simeq T_{0}-30\,\rm K$.\cite{As91,Rei95,Ga90}
Intriguing non--equilibrium behavior of segregation
amplitudes has been detected recently by time--resolved $X$--ray
experiments.\cite{Rei97} After a quench from an initial temperature
$T_{i}>T_{0}$ to a final temperature $T_{f}<T_{0}$ the initial surface
segregation profile induces a segregation front which rapidly progresses 
towards the bulk,
while lateral order was found to develop on longer time--scales. This
two--stage ordering process was interpreted recently in terms of
time--dependent Ginzburg--Landau (GL) theory.\cite{Fi99} Although 
restricted to one dimension and to the
limit of vanishing thermal noise, this theory successfully yields the
difference in time--scales for perpendicular and lateral ordering and
identifies an anisotropy of the evolving domain structure 
within a characteristic penetration depth of the segregation front.

Motivated by these studies of the subsurface ordering kinetics of
$\mathrm{Cu_{3}Au}$ (001), our aim here is to gain more insight into
possible surface--induced ordering scenarios in $A_{3}B$ fcc--alloys
with the help of dynamical Monte
Carlo simulations. As a minimal model with respect to equilibrium
properties we choose a lattice with nearest--neighbor interactions
among $A$ and $B$--atoms, known to account reasonably for the bulk
phase diagram of the $\mathrm{Cu-Au}$ system. Our dynamic algorithm relies on
the vacancy mechanism, where elementary atomic moves consist in an
exchange of $(A)\mathrm{Cu}$-- or $(B)\mathrm{Au}$--atoms with vacancies
$V$.\cite{Bi91a,Bi91b,Vi92,Fr94,Fra94,Gou97,Pu98,Po99} This kind of $ABV$--model, 
which has been used before by
Frontera et al. in an investigation of the bulk ordering kinetics in
$A_{3}B$ alloys,\cite{Fr97} is physically more realistic than
the conventional direct $AB$--exchange kinetics and will simultaneously
allow us to scale the simulated surface--induced growth rates with
the self--diffusion coefficient of $A$-- and $B$--atoms.

The main results of our simulations are: (i) the occurrence of two 
distinctly different modes of progression of the segregation front, 
depending on the final temperature $T_{f}$ to lie significantly above the
spinodal temperature, $T_{f}>T_{sp}$ (quench into the metastable
regime) or below, $T_{f}<T_{sp}$ (quench into the unstable regime), (ii) an 
early--time linear
(constant velocity) growth behavior of the segregation front in the case
$T_{sp}<T_{f}<T_{0}$ and a characteristic time $t^{*}$, where this
linear growth decelerates. Below $T_{sp}$ it changes over to a growth
roughly proportional to $t^{1/4}$ up to our longest computing times. 
(iii) A different behavior of lateral growth in the
near--surface region. In connection with this we give a
detailed description of the persistent anisotropy in the near--surface
domain--structure.

The paper is organized as follows.
Section 2 defines our model; its bulk properties and equilibrium
surface properties are
tested in Section 3. The subsequent Section 4 discusses in detail the
processes of perpendicular and lateral ordering. Our findings have
strong relevance to the temperature quench experiments on $\mathrm{Cu_{3}Au}$ 
described above, although the present model based on spatially uniform 
parameters for nearest--neighbor interactions cannot be expected to work
quantitatively. In fact it implies an enhancement of static lateral
order near the surface as compared to the bulk, an effect not borne out in
the real material $\mathrm{Cu_{3}Au}$. Enhanced lateral order at the 
(001)--surface, however, seems to be observable in the fcc--alloy 
$\mathrm{Cu_{3}Pd}$.\cite{Rei01} 
Concluding remarks are presented in Section 5.

\section*{2. Model and Simulation Method}

Consider a lattice of $N\times N\times M$ fcc-cells with periodic
boundary conditions in the $x$-- and $y$--direction. Along the
$z$--axis our system is bound between two free surfaces. Atomic layers
normal to the $z$--direction with distance $a$ are labeled by $n$ with $1\le n\le
2M+1$. Each lattice site $i$ can be occupied either by an $A$--atom, or a
$B$--atom, or a vacancy so that the corresponding occupation numbers
$c_{i}^{A}$, $c_{i}^{B}$ and $c_{i}^{V}$ satisfy
$c_{i}^{A}+c_{i}^{B}+c_{i}^{V}=1$. In our simplified alloy model only
pairwise nearest--neighbor interactions are taken into account. The
configurational Hamiltonian then reads
\begin{equation}\label{H}
H=\sum_{\langle
  i,j\rangle}\left[V_{AA}c_{i}^{A}c_{j}^{A}+V_{BB}c_{i}^{B}c_{j}^{B}+V_{AB}
  \left(c_{i}^{A}c_{j}^{B}+c_{i}^{B}c_{j}^{A}\right)\right],
\end{equation}
where the summation is over nearest-neighbor pairs.\cite{Blu71} In
view of the very small concentration of vacancies in real alloys \cite{Fu1} it is
natural to assume $c_{i}^{V}\ll 1$ such that macroscopic
{\em equilibrium} properties of the alloy phases will remain
essentially unaffected by the vacancies. For equilibrium
considerations one can then ignore the vacancies and introduce spin variables
$s_{i}=2c_{i}^{A}-1=\pm 1$ by which equation (\ref{H}) is mapped onto the 
spin--$1/2$ Ising model,
\begin{equation}\label{HI}
H_{I}=-J\sum_{\langle i,j\rangle}s_{i}s_{j}-h\sum_{i}s_{i}-h_{1}\sum_{i}\,'s_{i}
\end{equation}
with
\begin{equation}\label{J}
J=-\frac{1}{4}\left(V_{AA}+V_{BB}-2V_{AB}\right)
\end{equation}
\begin{equation}
h=3(V_{BB}-V_{AA})
\end{equation}
The last summation in (\ref{HI}) is restricted to spins in the surface layers 
$n=1$ and $n=2M+1$, which are subjected to the surface field
\begin{equation}\label{h1}
h_{1}=V_{AA}-V_{BB}
\end{equation}
Bulk equilibrium
properties at fixed composition only depend on the parameter $J$. Regarding the
Cu($A$)--Au($B$) system, it is natural to take $V_{BB}>0$
because of the larger size of $\mathrm{Au}$--atoms relative to the
$\mathrm{Cu}$--atoms, whereas $V_{AA}<0$, $V_{AB}<0$ such that bulk
interactions will favor ``anti-ferromagnetic'' ordering $J<0$. Near
the stoichiometric composition $A_{3}B$ the emerging bulk $L1_{2}$--structure is
described by a $4$--component order parameter, one conserved
concentration variable $\psi_{0}$ and three non-conserved components 
$\psi_{1},\psi_{2}$, and $\psi_{3}$. These
are defined via the differences
$m_{\alpha}=\langle c_{i}^{A}\rangle - \langle c_{i}^{B}\rangle;\;i\in\alpha;$ of
averaged $A-$ and $B-$occupations 
of the four equivalent
simple--cubic sublattices $\alpha=1,\ldots 4$ building the fcc
lattice, namely \cite{Dosch}
\begin{equation}\label{psi0}
\begin{split}
\psi_{0} & =m_{1}+m_{2}+m_{3}+m_{4}\\
\psi_{1} & =m_{1}-m_{2}-m_{3}+m_{4}\\
\psi_{2} & =m_{1}-m_{2}+m_{3}-m_{4}\\
\psi_{3} & =m_{1}+m_{2}-m_{3}-m_{4}\\
\end{split}
\end{equation}
Each non-conserved component describes a layerwise modulation of the
$B$--atom concentration along one of the three cubic axes. The ground
state shows $4$--fold degeneracy, where the four types of equivalent ordered 
domains have components 
$(\psi_{1},\psi_{2},\psi_{3})/\bar{\psi}=(-1,1,1);\,(1,-1,1);\,(1,1,-1)$
and $(-1,-1,-1)$. In the ground state ($T=0$) the amplitude
$\bar{\psi}$ takes the value $\bar{\psi}=2$. 

The above conditions for the interaction parameters imply a negative
surface field $h_{1}<0$, favoring an enrichment of $B$--atoms in the
outermost layers $n=1$ and $n=2M+1$. This is to be expected in view of
the $BB$--repulsion. As shown by Schweika and Landau \cite{Sch98} on the basis 
of the Hamiltonian (\ref{HI}) and single--spin flip dynamics, the value $h_{1}/|J|=-4$ is consistent with
the experimentally measured concentration of $\mathrm{Au}$--atoms of
about $50\%$ in the surface layer of $\mathrm{Cu_{3}Au(001)}$ near the
ordering temperature $T_{0}$.\cite{Rei95}

In our present work, all investigations will be based on
the atom--vacancy exchange mechanism, where the energetics are given
by the complete hamiltonian (\ref{H}). For the final choice of
interaction parameters see the next section. To affect an elementary
atomic move in the Monte Carlo runs we randomly select a
nearest--neighbor vacancy--atom pair and calculate the vacancy--atom
exchange probability by using the Metropolis algorithm. Throughout most
of our calculations the mean vacancy density is taken as $c^{V}\simeq 6.1\cdot
10^{-5}$, which amounts to $128$ vacancies in a system of typical size
$N=64;\,M=128$ with $4N^{2}M$ sites. One Monte Carlo step (MCS)
consists of $4N^{2}M$ attempted moves and is independent of $c^{V}$. Clearly,
in an actual material the vacancy concentration generally will depend
on temperature and the amount of $A$-- and $B$--atoms. Our kinetic
model with fixed $c^{V}$ therefore does not allow us to compare
time--scales of vacancy--mediated processes taking place at different 
temperatures and
composition. However, our subsequent studies of the ordering kinetics
focuses on the exact stoichiometric composition $A_{3}B$
($c^{B}=0.25$) and a narrow temperature range near $T_{0}$ where
changes of $c^{V}$ are expected to be of minor importance. Local order
parameters to be used in the analysis of our simulations are defined
in terms of averages of (\ref{psi0}) over one fcc--cell. Order
parameter profiles and structure factors are obtained by averaging
over 10 independent simulation runs. 

\section*{3. Bulk properties and surface segregation}

\begin{figure}
\begin{center}
\epsfig{file=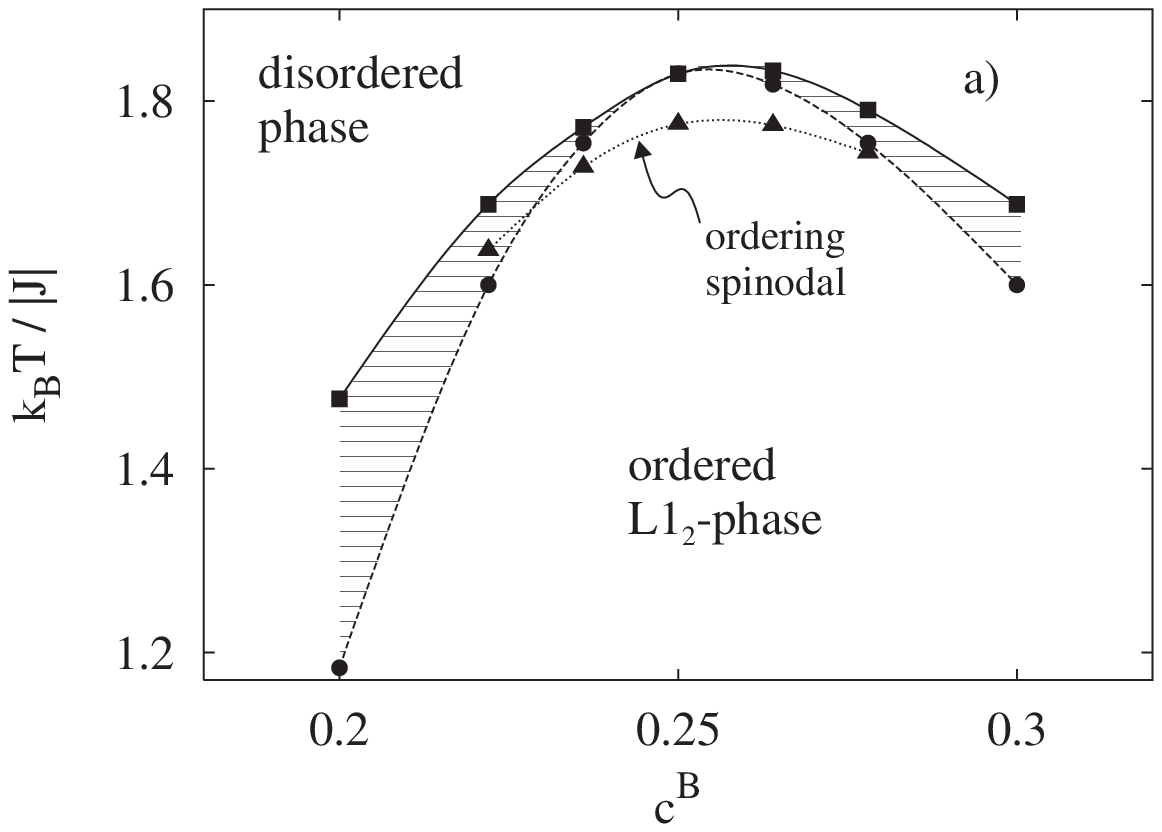, width=8cm}
\epsfig{file=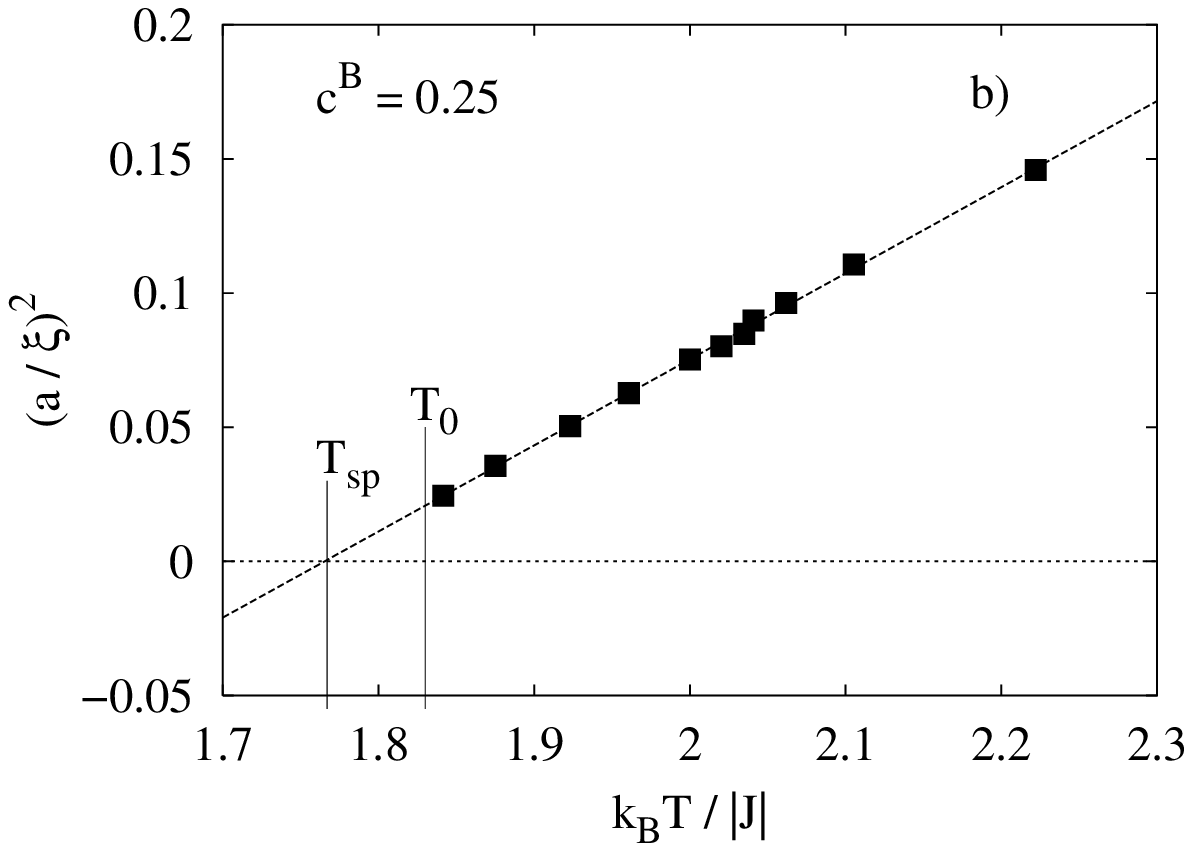, width=8cm}
\end{center}
\caption{(a) Phase diagram of the $AB$--binary alloy model in the
vicinity of $c^{B}=0.25$. Data points from simulations are
connected by smooth lines. Full line ($\blacksquare$) and dashed line 
($\bullet$): boundaries
of the disordered phase and the ordered $L1_{2}$--structure, with
the two--phase coexistence region in between. Dotted line 
($\blacktriangle$): ordering spinodal.\\
(b) Plot of $\xi^{-2}$ versus temperature at $c^{B}=0.25$,
illustrating the determination of the spinodal temperature.}
\label{Fig1}
\end{figure}

In this section we discuss some elementary properties of our model,
part of which are already known in the literature but they are needed here
as a test of our kinetic algorithm and as a basis for the subsequent
investigations of surface--induced kinetic properties. Fig.~\ref{Fig1}a shows
the phase diagram in the vicinity of $c^{B}\simeq 0.25$ as deduced
from simulations of a system with size $M=N=64$ and periodic
boundary conditions in all directions. To determine the ordering
temperature at stoichiometry ($c^{B}=0.25$), we analyzed the decay of a
perfectly ordered initial state upon thermalization at different
temperatures. This yields $k_{B}T_{0}\simeq 1.83|J|$, 
in good agreement with literature data
($k_{B}T_{0}\simeq 1.85|J|$).\cite{Fr97,Sch98,Bi80} From a similar
procedure, supplemented by computations of concentration 
($\psi_{0}$)--histograms, we
deduced estimates for the boundaries of the two--phase regions, which
reasonably reflect the bulk phase diagram of the
$\mathrm{Cu-Au}$ system near $c^{B}=0.25$. \cite{Fu2}

Next we proceed to equilibrium properties connected with the
($001$)--surface. From our algorithm defined in Section 2 we computed
equilibrated $B$--atom concentrations in the surface layers as a
function of $h_{1}$, keeping $J$ fixed. As before, $h_{1}$ and $J$ are
defined by (\ref{h1}) and (\ref{J}). Good agreement was found with the
work by Schweika and Landau.\cite{Sch98} We consider this both as a
positive test of our algorithm and as an indication that the small
amount of vacancies used has practically no
influence on these static results. In what follows we shall therefore
choose the interaction parameters of our $\mathrm{Cu_{3}Au}$--model
such that $h_{1}=-4|J|$.\cite{Sch98} While $J$ and $h_{1}$ are now
determined via the bulk ordering temperature $T_{0}$ and the
$\mathrm{Au}$--segregation amplitude in the surface layer one more
interaction parameter needs to be specified. Our final choice is
$V_{BB}=-V_{AA}=-V_{AB}>0$ with $|J|=V_{BB}/2$. This can be shown to
imply a weak effective attraction between vacancies.\cite{Po97,Po99}

Furthermore we remark that the
present model displays a surface transition at a critical temperature
$T_{cs}>T_{0}$, which we estimate as $T_{cs}\simeq 1.12\,T_{0}$ for
$c^{B}=0.25$. Below $T_{cs}$ the outermost layers with nearly equal
amounts of $A$-- and $B$--atoms develop a lateral
(1,1)--superstructure. With respect to $\mathrm{Cu_{3}Au}$, the
existence of a surface critical temperature clearly is an artefact of
the present model, connected with the fact that fcc
lattice models based on spatially uniform short range interactions are unable to
account for surface--induced disordering.\cite{Sch98} The (001)--surface of 
the $L1_{2}$--structure of
$\mathrm{Cu_{3}Pd}$, on the other hand, appears to favor lateral 
ordering.\cite{Rei01}

Moreover we estimate the spinodal temperature in a manner as done
experimentally.\cite{Rei95} In the disordered phase $B$--atoms are known to
show an oscillatory segregation profile near the (001) surface, with
enrichment and depletion in odd and even layers, respectively. When
atoms in the outermost layers are identified with sublattices
$\alpha=3$ and $4$, the segregation amplitude is given by the order
parameter component $\psi_{3}$. At the surface, its magnitude is
determined by the surface field $h_{1}$. Below the surface it decays
towards the bulk on a length scale given by the bulk correlation
length $\xi(T)$.\cite{Me95} Extrapolation of data for $\xi(T)$ to lower
temperatures according to $\xi(T)\sim(T-T_{sp})^{-1/2}$ allows one to
estimate the spinodal temperature $T_{sp}$. Experimentally,
$T_{0}\simeq 663\,K$ and $T_{sp}\simeq T_{0}-30\,K$ near the
stoichiometric composition of $\mathrm{Cu_{3}Au}$.\cite{Rei95} Application
of this procedure to our simulation data is illustrated in Fig.~\ref{Fig1}b and yields
the ordering spinodal displayed in Fig.~\ref{Fig1}a. At $c^{B}=0.25$
we find $T_{sp}/T_{0}\simeq 0.967\pm 0.003$.\cite{Funo} This is somewhat 
larger than the experimental value but reasonably close in view of the 
simplicity of our model.  
Simultaneously we get (see Fig.~\ref{Fig1}b)
$\xi(T_{0})\simeq 6a$, which again agrees fairly well with 
experiment.\cite{Rei95} 

\begin{figure}
\begin{center}
\epsfig{file=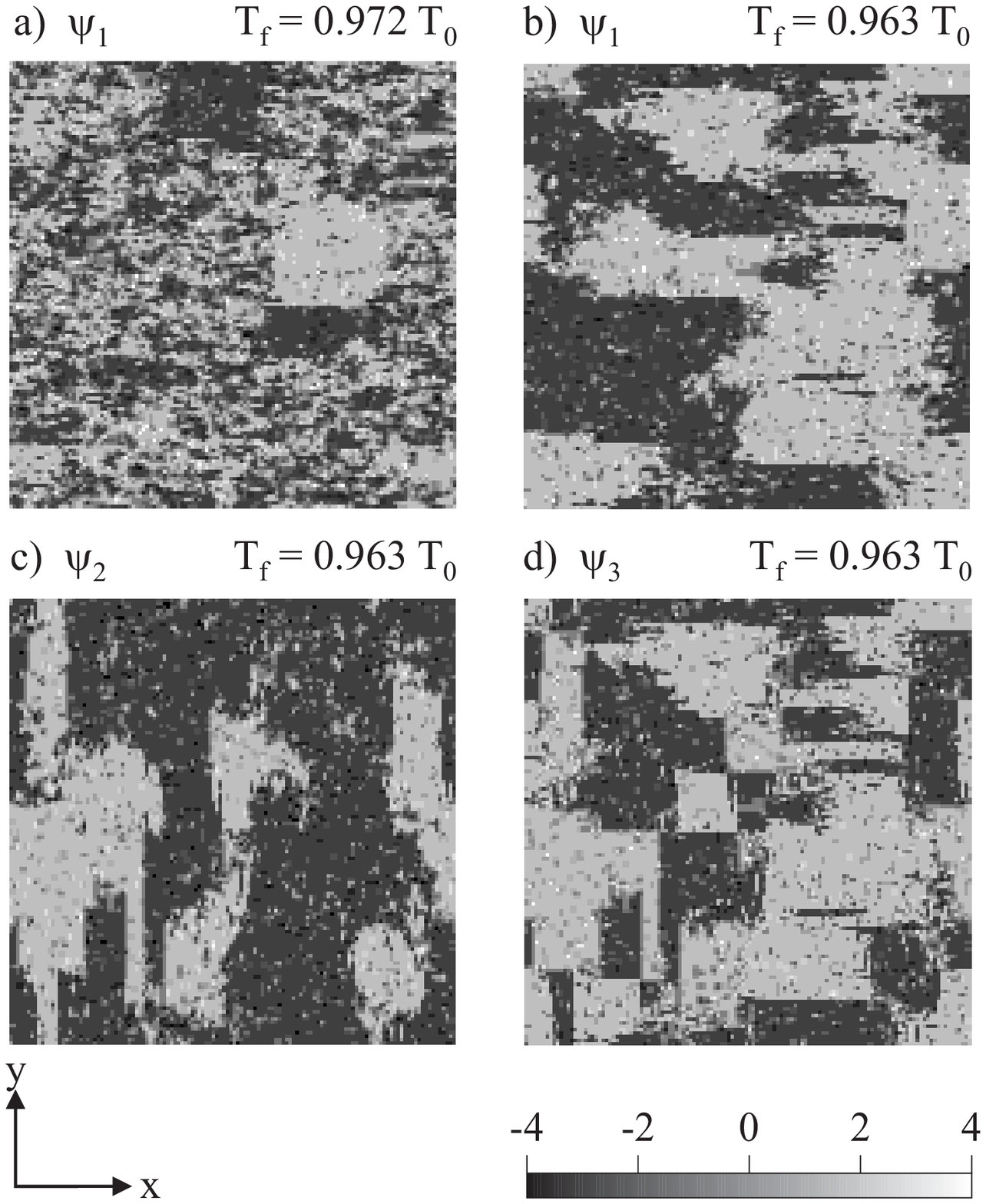, width=8cm}
\end{center}
\caption{Bulk domain patterns ($c^{B}=0.25$) at a time $t=7\cdot
10^{3}$ MCS after a quench from random initial conditions. Grey
scales reflect values of the local order parameters between extremal
values $4$ and $-4$.\\
(a) $\psi_{1}$--pattern at a final temperature
$T_{f}=0.972\,T_{0}>T_{sp}$.\\
(b)-(d) patterns of $\psi_{1},\psi_{2},\psi_{3}$ at
$T_{f}=0.963$, making evident the two types of domain walls and the
anisotropy of the evolving structures.}
\label{Fig2}
\end{figure}

For metallic alloys where short range
interactions prevail, it is well known that the spinodal and the
nucleation regime connect smoothly.\cite{Bi87} The qualitative
significance of the spinodal temperature, however, becomes apparent
from Fig.~\ref{Fig2}a and b.
There we compare bulk patterns of the order
parameter component $\psi_{1}$ at a time $t=7\cdot10^{3}$ MCS after a
sudden quench from a purely random initial state to final
temperatures $T_{f}\simeq 0.972\,T_{0}>T_{sp}$ (Fig.~\ref{Fig2}a) and
$T_{f}=0.963\,T_{0}\,\lsim \,T_{sp}$ (Fig.~\ref{Fig2}b), respectively. Snapshots
were taken in a section parallel to the ($xy$)--plane. 
In the first case 
only small ordered
domains have nucleated in an essentially disordered background
(Fig.~\ref{Fig2}a). By contrast in Fig.~\ref{Fig2}b we observe a continuous domain
pattern typical of spinodal ordering. At that temperature
($T_{f}=0.963\,T_{0}$) the same type of patterns is seen also at
shorter observation times. For a complete characterization
of the evolving domains in Fig.~\ref{Fig2}b we need in addition the
$\psi_{2}$-- and $\psi_{3}$--patterns displayed in Figs.~\ref{Fig2}c and
\ref{Fig2}d. From the grey--scales in Figs.~\ref{Fig2}b--\ref{Fig2}d one 
can easily see that inside a domain we have
$\psi_{1}\psi_{2}\psi_{3}<0$. Furthermore, the well--known two types of
antiphase boundaries \cite{Lai90} are recovered. For example, boundaries 
where $\psi_{1}$
and $\psi_{3}$ simultaneously change sign in the $y$--direction are the
``low--energy'' (type--1) domain walls and in fact cost no energy in our
nearest--neighbor model. These walls are almost completely flat and
turn out to be rather stable. On the other hand, boundaries where
$\psi_{1}$ and $\psi_{3}$ simultaneously change sign in the
$x$--direction are ``high--energy'' (type--2) walls where
curvature--driven coarsening is much more effective than for type--1
walls. Bulk coarsening effects at quench temperatures lower than those
considered here have been simulated in detail by
Frontera et al.\cite{Fr97} These authors obtained coarsening exponents $n$ 
smaller than that in the conventional
Lifschitz--Allen--Cahn coarsening law, where $n=1/2$, and interpreted their 
findings in terms of the presence of those stable low--energy walls.

\begin{figure}
\begin{center}
\epsfig{file=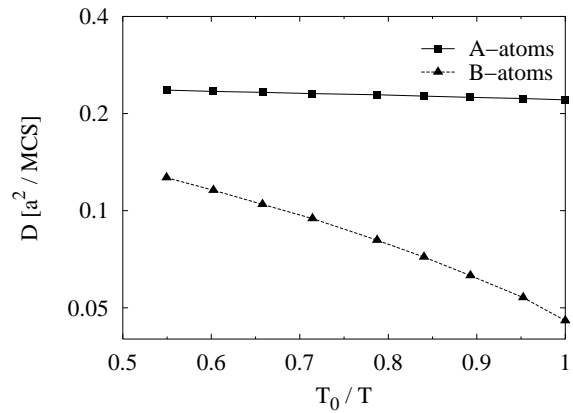, width=8cm}
\end{center}
\caption{Temperature--dependent diffusion constants of $A$-- and
  $B$--atoms in an Arrhenius representation.}
\label{Fig3}
\end{figure}

In addition we extract tracer diffusion coefficients from our model,
restricting ourselves again to $c^{B}=0.25$ and to temperatures above but
close to $T_{0}$. Diffusion coefficients $D_{A}$ and $D_{B}$ for
$A$-- and $B$--atoms are obtained in the usual way from
time--dependent mean--square displacements and are plotted in Fig.~\ref{Fig3}
in an Arrhenius representation. Since in our model $V_{AA}=V_{AB}$, the
energy remains essentially unchanged by a jump of an $A$--atom so that
$D_{A}$ is found to be nearly temperature--independent. On the other
hand, because of the
$B$--$B$--repulsion the $B$--atoms will migrate more slowly
than the $A$--atoms, and $D_{B}$ decreases upon cooling. As
indicated already, in a real system $c^{V}$ will grow with temperature. 
Therefore the individual experimental curves $\log D_{A}$
or $\log D_{B}$ versus $1/T$ will be steeper than those in
Fig.~\ref{Fig3}. However, the ratio $D_{A}/D_{B}$ roughly agrees with
experimental data for $\mathrm{Cu_{3}Au}$, where $D_{A}/D_{B}\simeq
1.45$ at $T/T_{0}\simeq 1.7$,\cite{Heu78} while we find $D_{A}/D_{B}\simeq 2$ 
at the same $T/T_{0}$--ratio.

Let us close this section by a brief remark on the distribution of
vacancies in our system with free surfaces. First of all, vacancies
tend to enrich in the outermost layers. For example at $T/T_{0}=0.972$
we find a surface vacancy concentration of about
$c^{V}_{\mathrm{surface}}\simeq 2.5\cdot 10^{-3}$, much larger than the 
overall concentration. The fact that vacancies are expelled from the bulk
suggests that in fact we are dealing with a model for a stable solid
which, when regarding the complete ternary ($ABV$)--system, would be
phase--separated from the ``vacuum'' (vacancy--rich)
phase. Furthermore, when analyzing the
above--mentioned layered structure along the $z$--axis caused by
surface segregation, we find that vacancies deplete in the $AB$--layers and
enrich in the $A$--layers. The reason is that the $BB$ repulsive energy in 
$AB$--layers becomes minimized when vacancies avoid $B$--sites. The resulting 
oscillations in the vacancy--density along the $z$--axis have an amplitude of 
about $30$ percent relative to the average vacancy density.

\section*{4. Surface--induced ordering kinetics}

\begin{figure}
\begin{center}
\epsfig{file=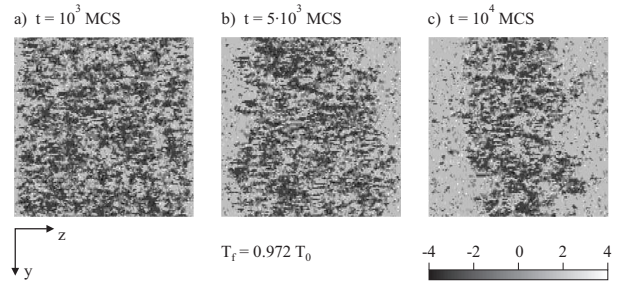, width=8cm}
\end{center}
\caption{$\psi_{3}$--patterns at $T_{f}=0.972\,T_{0}$ for three
different times after the quench, illustrating propagation of a
surface--induced ordering wave.}
\label{Fig4}
\end{figure}

\begin{figure}
\begin{center}
\epsfig{file=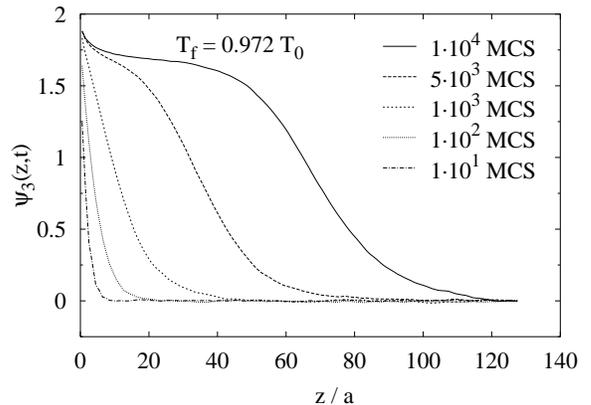, width=8cm}
\end{center}
\caption{Average profiles $\psi_{3}(z,t)$ for several times after
the quench.}
\label{Fig5}
\end{figure}

Now we turn to the question of how order evolves in a system with free
surfaces. Quench conditions are chosen as in Section 3 with final
temperatures below $T_{0}$ but in the vicinity of $T_{sp}$ ($0.9\lsim
T_{f}/T_{0}\lsim 1$). First, within few MC--steps $B$--atoms enrich in
the surface layer to a concentration of about $50\%$ and deplete in the
adjacent layer, thereby establishing local equilibrium as enforced by
the surface field $h_{1}$. This implies $\psi_{3}\simeq 1.85$ to
about $1.9$ at $z=0$ in the temperature interval
considered. This process in turn induces moves of $B$--atoms from even
to odd layers within an increasing depth from the surface, thus
forming a $\psi_{3}$--front which penetrates into the bulk. 
As an
example, we show in Fig.~\ref{Fig4} the $\psi_{3}$--distribution in a section
parallel to the ($y,z$)--plane at $T_{f}=0.972\,T_{0}>T_{sp}$ for
three different times after the quench. The patterns clearly
demonstrate a systematic progression of an ordered state with
fixed (positive) sign of $\psi_{3}$. 
Taking averages over sheets
parallel to the surface, we obtain profiles $\psi_{3}(z,t)$ which are plotted in
Fig.~\ref{Fig5}. 

At $t=10^{4}$ a plateau region for smaller $z$ has evolved
with $\psi_{3}$--values in the vicinity of the equilibrium
order parameter $\bar{\psi}\simeq 1.79$ at that temperature, before the
profile decays to zero for larger $z$.
Evidently, the thickness of the interface between the partially
ordered state and the bulk (not penetrated by the $\psi_{3}$--front)
increases with time, an effect expected to arise both from growing
intrinsic capillary wave fluctuations \cite{Row} and from bulk
fluctuations in all three order parameter components with the
character as in Fig.~\ref{Fig2}a, to be ``incorporated''
into the front during propagation.\\

\begin{figure}
\begin{center}
\vspace{0.5cm}
\epsfig{file=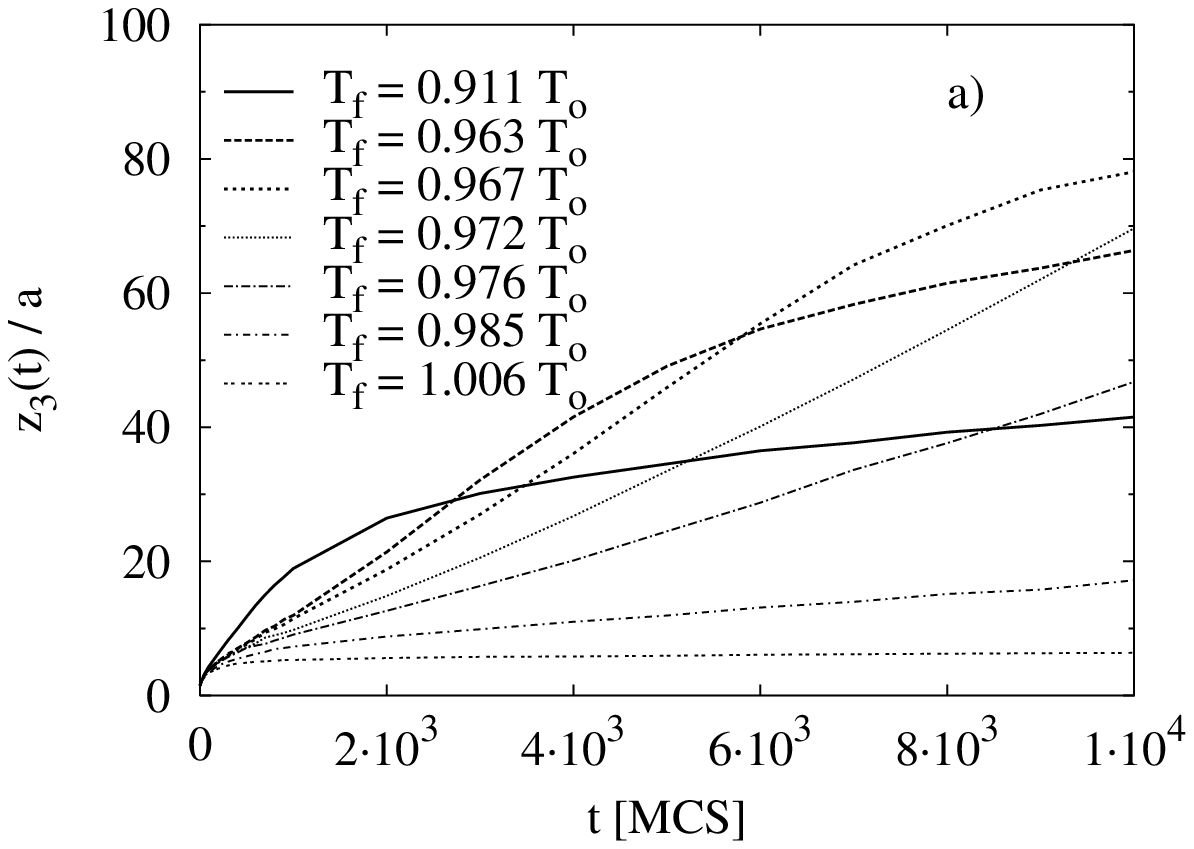, width=8cm}
\epsfig{file=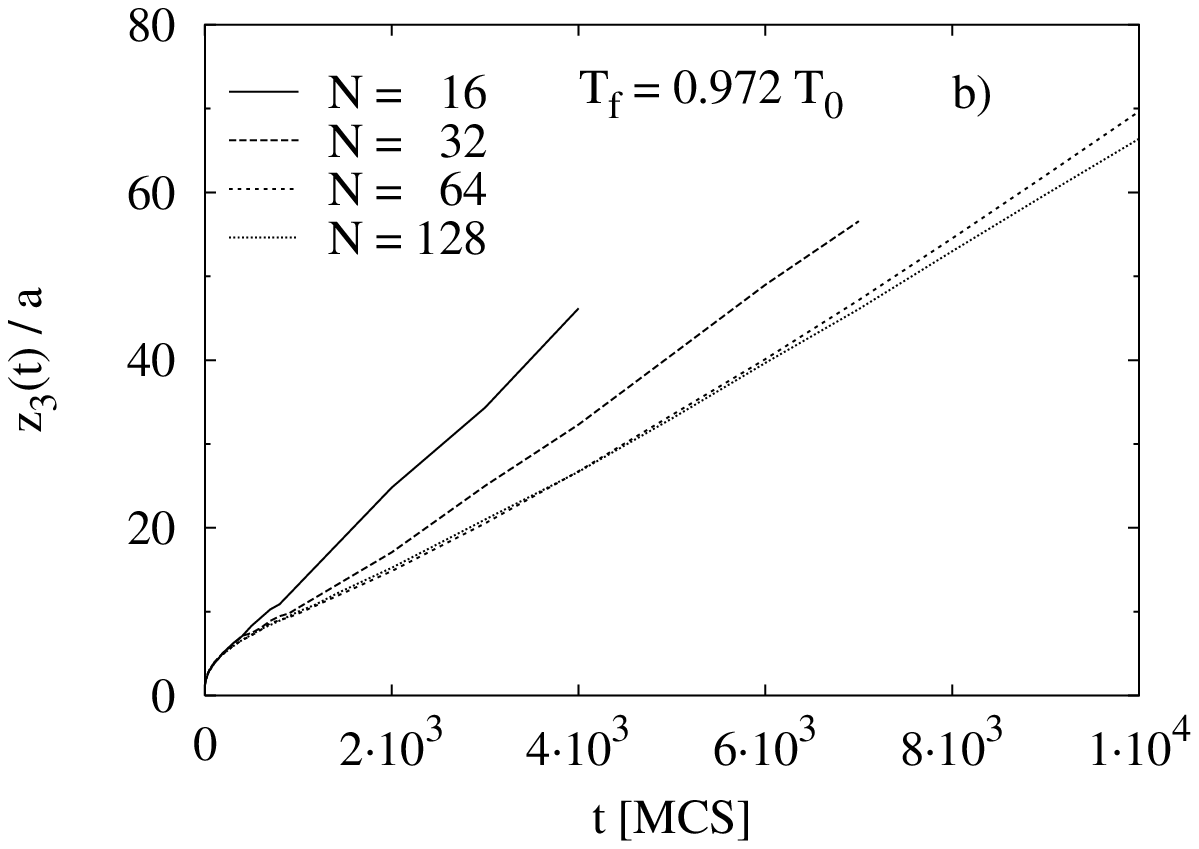, width=8cm}
\epsfig{file=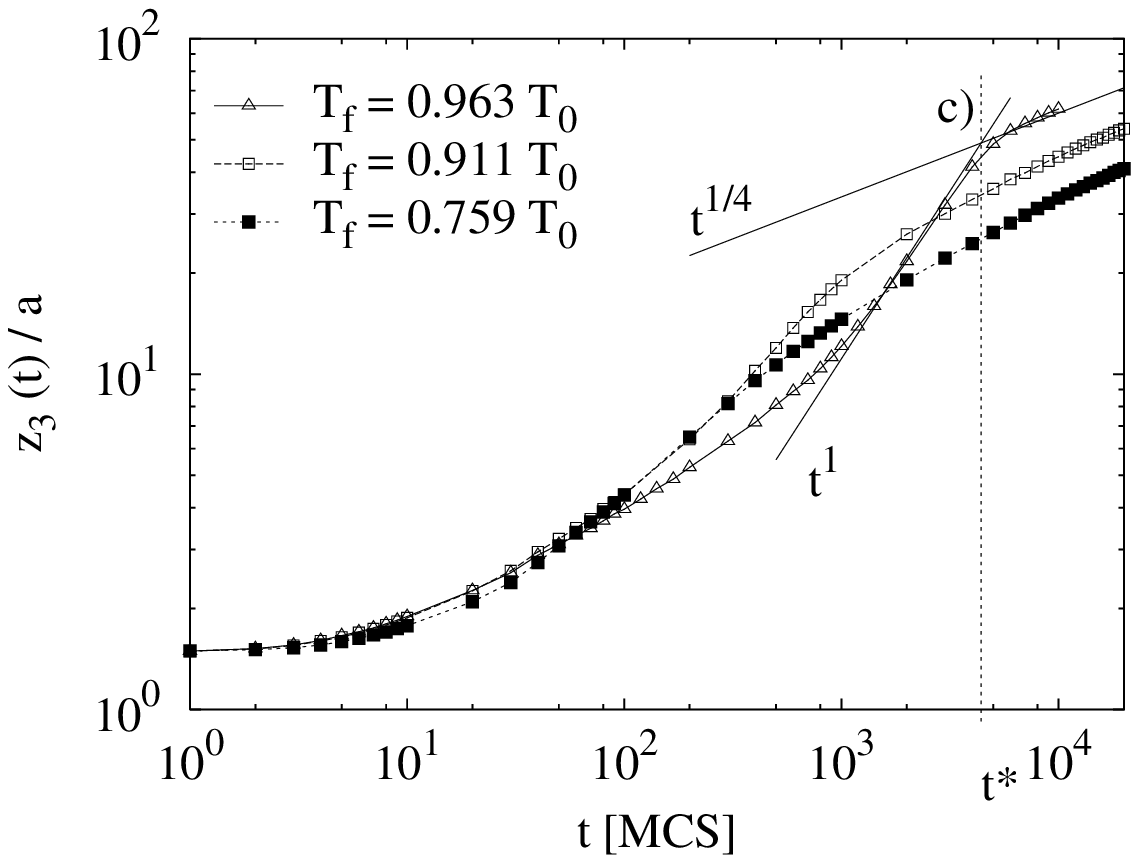, width=8cm}
\end{center}
\caption{(a) Time--dependent penetration depth $z_{3}(t)$ of the
$\psi_{3}$--front (see Fig. \ref{Fig5}) for various final temperatures.\\
(b) Dependence of $z_{3}(t)$ on the lateral system size $N$.\\
(c) Time--dependent penetration depth in a double--logarith\-mic
representation for three different temperatures. The growth behavior
for large times (up to $t=2\cdot 10^{4}$ MCS for the two lower
temperatures) is compared with a
$t^{1/4}$--law (straight line). At $T_{f}=0.963\,T_{0}$ a
comparatively short time domain of linear growth appears (see the
steeper straight line) up to the crossover time $t^{*}$.}
\label{Fig6}
\end{figure}

\begin{figure}
\vspace{0.0cm}
\begin{center}
\epsfig{file=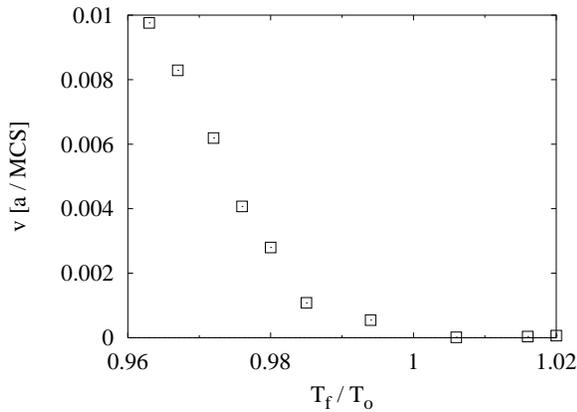, width=8cm}
\end{center}
\caption{Penetration velocity $v(T)$ versus temperature.}
\label{Fig7}
\end{figure}

The profiles $\psi_{3}(z,t)$ allow us to define a time--dependent
penetration depth $z_{3}(t)$ by setting
$\psi_{3}(z_{3}(t),t)=\frac{1}{2}\psi_{3}(0,t)$. Plots of $z_{3}(t)$
for a series of final temperatures $T_{f}$ are shown in Fig.~\ref{Fig6}a for a
system large enough in the $z$--direction ($N=64,\,M=128$) to
avoid the $\psi_{3}$--fronts induced by both surfaces to overlap. (In
fact, data in Fig.~\ref{Fig5} and \ref{Fig6} contain an average over those two fronts.)
As expected, after a quench not passing the phase boundary $z_{3}(t)$
rapidly saturates. For example, at $T_{f}=1.006\,T_{0}$ the final
penetration depth is about 6 atomic layers, which is of the order of
the bulk correlation length $\xi$ in the disordered phase at that
temperature. On the other hand, when $T_{sp}<T_{f}<T_{0}$, we see that
after a short transient $z_{3}(t)$ grows more or less
linearly with time within an extended time interval. This allows us to
introduce a penetration velocity $v(T)$ which increases upon lowering
the temperature. For example, at $T_{f}=0.972\,T_{0}$ we find that
$z_{3}(t)$ grows linearly up to at least $10^{4}$ MCS, where it reaches
about $70$ atomic layers, see also Fig.~\ref{Fig4}. In determining $v(T)$,
shown in Fig.~\ref{Fig7}, care has to be taken with respect to the lateral size $N$ of
the system. For small $N$ the system is likely to develop single
(closed) domains in the lateral direction within the front
region. Increased order of this type will speed up the penetration
process. This $N$--dependence of $v$ is exemplified in Fig.~\ref{Fig6}b for 
$T_{f}=0.972\,T_{0}$. Data up to $t=10^{4}$ MCS,
however, seem to converge sufficiently at our largest
$N=128$. Moreover, using again $N=64$, we have also tested the
influence of the vacancy concentration on $v$. At
$T_{f}=0.972\,T_{0}$, results with $c^{V}=3.05\cdot 10^{-5}$ ($64$
vacancies) are indistinguishable from those shown in the figure. Hence
the vacancy system appears to be in the dilute limit also with respect
to growth processes of this type. Note that for $\mathrm{Cu_{3}Au}$
the experimental vacancy concentration at $T_{0}$ is even lower.\cite{Fu1}

As seen
from Fig.~\ref{Fig7}, when temperature is decreased from $T_{0}$, the velocity
$v(t)$ first increases with upward curvature, but decelerates when
$T\approx T_{sp}$. Sufficiently below $T_{sp}$, linear growth can no
longer be identified and $v(T)$ becomes
undefined, see the case $T_{f}=0.911\,T_{0}$ in Fig.~\ref{Fig6}a. Some
estimates based on mean--field arguments concerning the order of
magnitude of $v(T)$ are presented in the Appendix.

When temperature falls near or below $T_{f}\simeq 0.963\,T_{0}$, a new
regime of slower growth beyond some crossover time $t^{*}$ becomes
detectable within the time window of Fig.~\ref{Fig6}. $t^{*}$ decreases with
temperature, so that the regime of linear growth shrinks. Notice in 
addition the non--monotonous behavior of
$z_{3}(t)$ as a function of temperature for fixed observation time
$t$. As indicated already, the strong shrinkage of the linear regime, where
$t^{*}$ tends to merge with the short time transient, occurs at
temperatures slightly below but very near the spinodal
temperature $T_{sp}\simeq 0.967$ discussed in Section 3 on the basis
of static considerations. Regarding these surface phenomena we conclude that 
the spinodal temperature as estimated via the static
segregation profiles separates two distinct dynamical regimes,
roughly characterized by the presence or the absence of
surface--induced linear (constant velocity) growth in an intermediate time 
domain.

The slowing down of the front motion beyond the linear regime,
i.\,e. for $t>t^{*}$, clearly originates from its competition with
nearly saturated domains that have already nucleated in the
bulk. Crudely speaking, $t^{*}$ will be determined by the
interplay of two
dynamic quantities, the penetration velocity $v(T)$ and the nucleation
rate of ordered domains. Moreover, unstable
growth of bulk domains at temperatures $T_{f}<T_{sp}$ totally
suppresses the linear regime. To investigate this situation more
closely, we carried out even longer runs up to $t=2\cdot 10^{4}$ for
the case $T_{f}=0.911\,T_{0}$. Results at that temperature for $z_{3}(t)$ in a 
log--log representation are shown in Fig.~\ref{Fig6}c. The data in the range
$t\gsim 2\cdot 10^{3}$ MCS appear to follow a power--law
$z_{3}(t)\simeq t^{n}$ with $n\simeq 0.27\pm 0.02$. 

The appearance of such 
small growth exponent is confirmed by additional simulations at a temperature
$T_{f}=0.759\,T_{0}$ considerably below $T_{sp}$, also shown in
Fig.~\ref{Fig6}c. These data indicate in addition that the onset of this type
of power law is moved to earlier times when temperature is
lowered. Conversely, data for $T_{f}=0.963\,T_{0}$ display a crossover
to a slow temporal growth compatible with $n\approx 1/4$ at a larger
crossover time, indicated as $t^{*}$ in Fig.~\ref{Fig6}c. Since the penetration depth 
up to $t=2\cdot 10^{4}$ stays significantly smaller than the system size 
(in this case $N=M=128$), we expect that finite size effects do not play an 
essential role in estimating that exponent $n$. Preliminary results for the
structure factor of a bulk system seem to confirm that value of $n$,
although no definitive conclusions
concerning the asymptotic long--time behavior can be
drawn from these studies so far. Nevertheless it is interesting to compare our 
findings in Fig.~\ref{Fig6}c with recent work by Cast\'an and 
Lindg\aa rd.\cite{Cas89,Cas91} These authors studied coarsening processes in a
$2$--dimensional system involving both curved and
flat (zero curvature) domain walls. The latter were essentially
immobile and could be removed only by the progression of an intervening
curved wall. Such a situation basically is encountered also in the
present model, where type--1 antiphase boundaries turn out to be
essentially immobile. This stability of flat walls induces an overall growth
exponent which is smaller than the classical Lifschitz--Allen--Cahn exponent
$n=1/2$.\cite{Lif62,Al79} Cast\'an in fact gave arguments in favor of an
exponent $n=1/4$. A similar conclusion with respect to $n$ was reached
by Deymier et al.\cite{Dey97} from studies of a different model with two types of contacts
between domains.

\begin{figure}
\vspace{-0.0cm}
\begin{center}
\epsfig{file=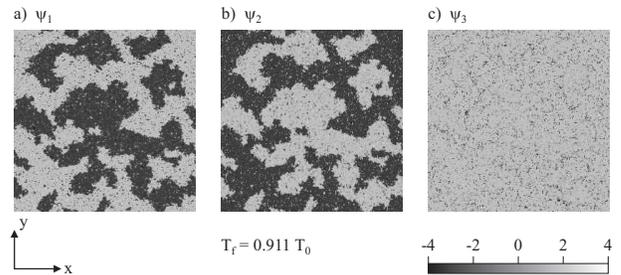, width=8cm}
\end{center}
\caption{Lateral domain patterns for a) $\psi_{1}$ and b) $\psi_{2}$ in
the near--surface penetrated region with nearly uniform $\psi_{3}$
(see c)) at $T_{f}=0.911\,T_{0}, t=10^{4}$ MCS. Notice the absence
of type--1 walls (see text).}
\label{Fig8}
\end{figure}

Next we study lateral ordering within layers parallel to the
($xy$)--plane at a depth $z$ from the surface. This process proceeds
in different ways depending on whether $z<z_{3}(t)$ or
$z>z_{3}(t)$. When analyzing the layer $z=M$ midway between the two
surfaces, representative for the latter case, we recover the
characteristics of bulk ordering as described briefly in Section 3,
see also Fig.~\ref{Fig2}. Our main objective, however, is the near--surface
region $z<z_{3}(t)$ and consequences of the surface--induced segregation wave. 
As emphasized before,\cite{Rei97,Fi99} for
$z<z_{3}(t)$ the sign of $\psi_{3}$ is essentially fixed to
$\psi_{3}>0$ so that only two types of domains can appear:
$(\psi_{1},\psi_{2},\psi_{3})/\bar{\psi}=(1,-1,1)$ and
$(-1,1,1)$. These correspond to the two ways $B$--atoms can
form a $(1,1)$--superstructure on the square lattice building an
odd ($AB$--type) layer. Walls between those domains encountered when
going in the lateral direction are exclusively of
type 2, whereas type--1 walls are parallel to the surface. To analyze
this situation further, we have enlarged the system in the lateral
direction to $N=256$, but restricted ourselves to $M=8$ so that
$\psi_{3}$ is nearly uniform along the $z$--axis. Patterns of all three
order parameter components were extracted from atomic layers $7$ and $8$ 
midway between the two surfaces. An example with $T_{f}=0.911\,T_{0}$ and 
$t=10^{4}$ MCS
is shown in Fig.~\ref{Fig8}. In fact, domain walls in Figs.~\ref{Fig8}a and \ref{Fig8}b with
simultaneous sign changes of $\psi_{1}$ and $\psi_{2}$ are always
curved, indicative of type--$2$ walls, and the domain structure appears
completely isotropic, in contrast to the anisotropic shape of bulk
domains (cf. Fig.~\ref{Fig2}). Fig.~\ref{Fig8}c depicts $\psi_{3}$ which is nearly
constant.

\begin{figure}
\begin{center}
\epsfig{file=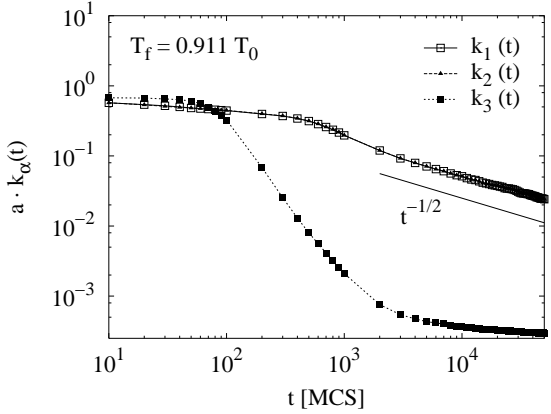, width=8cm}
\end{center}
\caption{First moments $k_{\alpha}(t)$ of lateral
structure factors $S_{\alpha}(k_{\parallel},z,t)$ in the same
($x,y$)--section as in Fig. \ref{Fig8}, for $T_{f}=0.911\,T_{0}$. The behavior of
$k_{1}(t)=k_{2}(t)$ at long times is compared with the 
Lifschitz--Allen--Cahn prediction (straight line).}
\label{Fig9}
\end{figure}

The time--dependent average size of domains in the ($xy$)--section of
Fig.~\ref{Fig8} is analyzed in Fig.~\ref{Fig9}, where we present results for the first
moment 
\begin{equation}
k_{\alpha}(z,t)=\frac{\sum_{k_{\parallel}}k_{\parallel}S_{\alpha}
(k_{\parallel},z,t)}{\sum_{k_{\parallel}}S_{\alpha}(k_{\parallel},z,t)}
\end{equation}
of the time--dependent lateral structure factor, defined by 
\begin{equation}
S_{\alpha}(k_{\parallel},z,t)=\langle|\psi_{\alpha}
(k_{\parallel},z,t)|^{2}\rangle
\end{equation}
where $\alpha=1$ to $3$ and $\psi_{\alpha}(k_{\parallel},z,t)$ is the
lateral Fourier transform of the local order parameter
$\psi_{\alpha}(\vec{r},t)$. Note that vacancy depletion in $AB$--type layers 
and enrichment in $A$--type layers (see Section 3) expands the overall
time--scale for lateral ordering. While the behavior of $k_{3}(t)$
reflects a rapid transition to a near--uniform
$\psi_{3}$--distribution (see Fig.~\ref{Fig8}c), the data points for $k_{\alpha}(t)$
with $\alpha = 1,2$
in Fig.~\ref{Fig9} seem to approach a growth behavior $k_{\alpha}^{-1}(t)\sim t^n$ 
described by the Lifschitz--Allen--Cahn exponent $n=1/2$. In fact, our 
results indicate curvature driven coarsening in the lateral directions, 
distinctly different from the perpendicular growth displayed in Fig.~\ref{Fig6}c.

\section*{5. Summary and concluding remarks}

Surface--induced ordering processes in an fcc alloy model were studied
on the basis of the atom--vacancy exchange mechanism. One essential result 
of our simulation studies is the appearance
of a segregation wave induced at the surface and propagating into the
bulk immediately after a temperature quench across the ordering
temperature $T_{0}$. This feature of our model qualitatively agrees
with peculiar subsurface ordering phenomena at
$\mathrm{Cu_{3}Au}$ (001) observed in recent $X$--ray experiments.\cite{Rei97} 
In the region $z<z_{3}(t)$ covered
by the segregation wave, the present model with 
nearest--neighbor interactions only shows an increased tendency to
build up lateral order. The experimental verification of this effect
may depend on the alloy material under consideration.

A second important conclusion confirmed by our simulations is the
existence of an anisotropic domain structure in the near--surface
region $z<z_{3}(t)$. Only two types of domains and, for $z$
fixed, only high--energy domain walls appear, as opposed to the
well--known domain structure in the bulk with $4$ types of domains and
both high-- and low--energy walls in either direction. Since in
$\mathrm{Cu_{3}Au}$ $z_{3}(t)$ can become larger than $200\,$\AA \cite{Rei97} 
or about $50$ atomic layers (in our simulations
more than $70$ atomic layers) this should offer the possibility to grow
films of mesoscopic thickness with an anisotropic domain structure not
realized in the bulk. An interesting question for future work concerns
films with odd and even numbers of atomic layers and overlapping
$\psi_{3}$--fronts induced by both surfaces.

Moreover, our present studies exemplify different ordering scenarios that may 
emerge more generally in systems whose bulk structure requires a multicomponent
order parameter, and where the surface favors some kind of partial
ordering. Concerning domain growth in our model, we have to distinguish between
perpendicular and lateral growth modes, between a near--surface
partially ordered region ($z<z_{3}(t)$) and the bulk
region ($z>z_{3}(t)$), and between quench temperatures $T_{f}$ above
and below the spinodal temperature $T_{sp}$. At least three distinct coarsening 
schemes at intermediate or long times were observed. 
If $T_{sp}<T_{f}<T_{0}$, perpendicular order initially grows linearly with $t$ 
such that $z_{3}(t)\simeq vt$ for $t<t^{*}$, while for $t>t^{*}$
our data indicate a temporal regime where $z_{3}(t)\sim t^{1/4}$. 
On the other hand, for $T_{f}<T_{sp}$ the regime of linear
perpendicular growth is suppressed. Analysis of the typical domain structure
in the lateral direction shows that in the region $z<z_{3}(t)$ it
is isotropic and coarsens according to $k_{\alpha}^{-1}(t)\sim
t^{1/2}$ at long times ($\alpha=1;2$) while for $z>z_{3}(t)$ one
recovers bulk
behavior. 

Besides some open questions addressed already above we like to point
out that non-stoichiometric alloys should in principle display even
more rich and interesting behaviors, in particular in cases where
ordering and phase separation occur simultaneously.\cite{Fi97}

\section*{Acknowledgments}

The authors are indebted to H. Reichert for several discussions
concerning the experimental situation and to P. Maass for helpful
comments. A useful conversation with M. Maret, M. Albrecht and
G. Schatz about surface segregation effects in the fcc alloy
$\mathrm{CoPt_{3}}$ is also gratefully acknowledged. This work was
supported in part by the Deutsche Forschungsgemeinschaft, SFB 513.

\section*{Appendix}

We attempt here to give some qualitative estimates concerning the magnitude 
of the penetration velocity
$v(T)$, see Section 4, from simplified mean--field considerations. Our
starting point are time--dependent Ginzburg--Landau equations, based
on the free energy density $f(\psi_{1},\psi_{2},\psi_{3})$ as given by
Lai \cite{Lai90} for the $\mathrm{Cu_{3}Au}$ structure,
\begin{equation}\label{f}
f=\sum_{\alpha}\left(\frac{r}{2}\psi_{\alpha}^{2}+
\frac{v}{4}\psi_{\alpha}^{4}\right)+\frac{u}{4}
\left(\sum_{\alpha}\psi_{\alpha}^{2}\right)^{2}+w\psi_{1}\psi_{2}\psi_{3}
\end{equation}
with $r=r_{0}(T-T_{sp});r_{0}>0;u>0;|v|<u;w>0$. The ordering
temperature $T_{0}$ is determined by $r(T_{0})=(2/9)w^{2}/(3u+v)$ and
the order parameter at $T_{0}$ has the magnitude
$|\psi_{\alpha}|=\bar{\psi}$ with $\bar{\psi}=(2/3)w(3u+v)$. Regarding
the penetration of the segregation wave as a one--dimensional process
along the $z$--axis (perpendicular to the surface), we assume an
equation of motion for $\psi_{3}(z,t)$ of the form
\begin{equation}\label{pt}
\partial_{t}\psi_{3}=-\Gamma\left(\frac{\partial
    f}{\partial\psi_{3}}-K\psi_{3}''\right)
\end{equation}
with a kinetic coefficient $\Gamma$ and $K>0$. Two limiting cases 
concerning the coupling of $\psi_{3}$
with $\psi_{1}$ and $\psi_{2}$ can be considered such that (\ref{pt})
becomes a closed equation for $\psi_{3}$ only.
Thereby it turns out that the simulated velocity falls between these
two limits. First, if the
parameters are such that the evolution of lateral order,
expressed by $\psi_{1}$ and $\psi_{2}$, considerably lags behind the
progression of the $\psi_{3}$--front, one may set in (\ref{pt})
$\psi_{1}=\psi_{2}=0$. In this case, the form (\ref{f}) of the free
energy density allows a nontrivial solution for a propagating front
only at temperatures $T<T_{sp}$, with a velocity of propagation
$v\propto(T_{sp}-T)^{1/2}$.\cite{Fi99} Correspondingly, we obtain $v=0$
for $T>T_{sp}$, contrary to Fig.~\ref{Fig7}. However, simulations of our
nearest--neighbor interaction model indicate that lateral order at
least on
small length scales evolves almost simultaneously with the front motion
in the perpendicular direction. This suggests to consider a second, in
fact opposite limit, namely that $\psi_{1}$ and $\psi_{2}$
instantaneously relax towards the ``local equilibrium condition''
$\psi_{1}=-\psi_{2}=\pm\psi_{3}$. This yields the approximation
$w\psi_{1}\psi_{2}\simeq -w\psi_{3}^{2}$, to be used in
(\ref{pt}). Introducing $\varphi=\psi_{3}/\bar{\psi}$ and the ``potential''
\begin{eqnarray}
V(\varphi) & = & -f/(3r(T_{0})\bar{\psi}^{2})=\nonumber\\
& = &
\frac{1}{2}\left(1-r/r(T_{0})\right)\varphi^{2}-\frac{1}{2}
\varphi^{2}(\varphi-1)^{2}
\end{eqnarray}
we obtain
\begin{equation}
\partial_{t}\varphi=\Gamma
r(T_{0})(dV/d\varphi+(\xi(T_{0}))^{2}\varphi'')
\end{equation}
where $\xi_{0}=(K/r(T_{0}))^{1/2}$ denotes the correlation length at
$T_{0}$. The ansatz $\varphi(z,t)=\eta(z-vt)$ leads to an ordinary
differential equation for $\eta(z)$. The velocity $v$ is determined in
the usual manner \cite{La83} by requiring that appropriate boundary conditions
can be fulfilled. These are $\eta(z)\rightarrow\eta_{\mathrm{max}}$;
$\eta'(z)\rightarrow 0$ as $z\rightarrow -\infty$, where
$\eta_{\mathrm{max}}>0$ corresponds to the absolute maximum of $V(\eta)$ and
$\eta(z)\rightarrow 0$; $\eta'(z)\rightarrow 0$ as
$z\rightarrow\infty$. Numerical solution of this problem shows that, at
$T_{0}$, $v(T)$ starts to grow linearly in $T_{0}-T$ and that
$v(T_{sp})\simeq 1.4\Gamma r(T_{0})\xi_{0}$. To estimate $\Gamma$,
one can formulate mean--field arguments for the diffusion of the slower
atomic species, in this case the $B$--atoms, along the $z$--axis,
which yields $D_{B}/a^{2}\simeq 2\Gamma r(T_{0})$. Using
$\xi_{0}\simeq 6a$, this gives the relation $v(T_{sp})\simeq
4(D_{B}/a)$, independent of the Monte Carlo time unit. From the
simulated value $D_{B}\simeq 0.04 a^{2}/\mathrm{MCS}$ at $T_{0}$, see Fig.~\ref{Fig3},
we obtain $v(T_{sp})\simeq 0.16a/\mathrm{MCS}$. Comparison
with Fig.~\ref{Fig7} shows that these simple arguments overestimate the
velocity at $T_{sp}$ by about one order of magnitude. This is not
surprising in view of the simplifications made, especially in view of
the neglect of lateral fluctuations. These may strongly reduce the penetration 
velocity, an effect already inferred
from the $N$--dependence of the velocity displayed in Fig.~\ref{Fig6}b.

\end{multicols}

\begin{thebibliography}{ccccc}
\bibitem{Funo} In a good approximation the main decay of the
  segregation profile towards the bulk follows an exponential,
  $\psi_{3}\propto\exp(-z/\xi(T))$, as long as $T>T_{cs}$. In the case 
  $T<T_{cs}$, $\xi(T)$ was extracted from the tails of the $\psi_{3}$--profiles.
\end{thebibliography}
\end{document}